\documentclass[prd,superscriptaddress,nofootinbib,amsmath,amssymb,aps,11pt]{revtex4-2}
\usepackage[utf8]{inputenc}

\usepackage{bm}
\usepackage{amsfonts}
\usepackage{latexsym}
\usepackage{graphicx}
\usepackage{amsmath}
\usepackage{anyfontsize}
\usepackage{palatino}
\usepackage{mathpazo}
\usepackage{bigints}
\usepackage[normalem]{ulem}
\usepackage{soul}
\usepackage{comment}

\usepackage{xcolor}     

\newcommand{\AP}[1]{\textcolor{blue}{#1}}

\begin{document}
\title{\boldmath Virial identities across the spacetime}

\author{Alexandre M. Pombo}
\email{pombo@fzu.cz}
\affiliation{CEICO, Institute of Physics of the Czech Academy of Sciences, Na Slovance 2, 182 21 Praha 8, Czechia}

\author{Lorenzo Pizzuti}
\email{lorenzo.pizzuti@unimib.it}
\affiliation{Dipartimento di Fisica G. Occhialini, Universit\'a degli Studi di Milano Bicocca, Piazza della Scienza 3, I-20126 Milano, Italy}

\begin{abstract}
    Virial-like identities obtained through Derrick's scaling argument are powerful, multi-purpose tools to study general relativistic models. Applications comprise establishing no-go/hair theorems and numerical accuracy tests. In the presence of a horizon (\textit{aka} boundary), the spacetime can be divided into regions, each with its own identity. So far, such identities have only been computed in the region outside the event horizon; however, adding a positive cosmological constant endows an additional boundary (the cosmological horizon), with the region between the latter and the former of particular interest. In this letter, by performing a radial coordinate transformation, we generalise Derrick's scaling argument to compute virial identities \textit{across the whole} non-asymptotically flat spacetimes. The developed method is applied to the entire Reissner-Nordstrom-de Sitter spacetime. A convenient gauge that trivialises the gravitational contribution to the identity between horizons is also found.
\end{abstract}

\maketitle

%
\section{Introduction}\label{S1}
%
     In a stable system, \textit{virial identities}~\cite{clausius1870xvi} relate the average (over time) kinetic energy with half of the average potential energy, resulting in a meaningful relation that has been extensively used in physics. Besides the traditional applications in classical mechanics, in statistical mechanics, the virial identity allows the calculation of the average kinetic energy (temperature) of very complex many-particle systems from the knowledge of their position (potential energy). In astrophysics, the virial theorem has been used first to infer the presence of additional matter components in clusters of galaxies~\cite{zwicky1979masses}, later called dark matter, and nowadays it is widely applied in the kinematic mass reconstruction of clusters (\textit{e.g.} \cite{Munari2013,Biviano2023}).

    In field theory, virial identities can be computed by scaling the radial coordinate. First proposed by G. Derrick~\cite{derrick1964comments}, such scaling approach allowed to establish a generic argument against the existence of stable, finite energy, time-independent solutions in a broad class of non-linear wave equations – see also~\cite{hobart1963instability,hobart1965non} and \cite{deser1976absence} for an earlier similar argument. 

    In previous papers~\cite{herdeiro2021virial,herdeiro2022deconstructing,oliveira2023convenient}, a generalised framework of virial identities for both spherical and axially symmetric, asymptotically flat spacetimes, in the region outside the event horizon was presented. As a summary, consider the following one-dimensional (1D) effective action (EA):
	\begin{equation}\label{E1}
	   \hat{\mathcal{S}}\big[ X(r),X'(r),X''(r),r\big ] = \int_{r_i}^{\infty}\hat{\mathcal{L}}\left(X,x',r\right)\text{d}r \ ,
	\end{equation}
	where $X\equiv X_j(r)$ are the generic set of solely "radial``-dependent functions that parameterise our model. Prime denotes derivative with respect to the radial coordinate, $X'\equiv \text{d}X/\text{d}r$, and $r_i$ is some appropriately chosen constant that defines the boundary at smallest radius: $r_i=0$ for solitons and $r_i=r_H$ is the event horizon radius for BHs\AP{\footnote{Examples of the former can be seen at \cite{herdeiro2017asymptotically,pombo2023coupled} while of the latter in \cite{pombo2023effects,oliveira2021spontaneous,fernandes2019spontaneous}}}. The \textit{effective Lagrangian} $\hat{\mathcal{L}}$ should depend on $X''$, which can be absorbed into a function $f$ of the same variables as the non-total derivative piece of the effective Lagrangian (\textit{aka} reduced Lagrangian), $\mathcal{L}$,
	\begin{equation}\label{E2}
         \hat{\mathcal{L}}\left(X,X',X'',r\right)=\mathcal{L}\left(X,X',r\right)+\frac{d}{dr} f \left(X,X',r\right) \ .
	\end{equation}
    In the presence of a boundary at $r_i$, Derrick's standard scaling argument, $r \to r_i + \nu(r-r_i)$, first translates the radial coordinate as $x \to (r-r_i)$ and then scales the radial coordinate as:
	\begin{equation}\label{E3}
	   x\to \nu x \ . 
	\end{equation}
	which induces a variation of any fiducial configuration $X(r)$, as $X(r)\rightarrow X(\nu r)=X_{\nu}(x)$. Observe that the radial translation of the event horizon to the origin of $x$ makes it invariant under the scaling \eqref{E3}, effectively "fixing´´ the horizon.

	The EA of the scaled configuration $X_{\nu}(r)$ becomes a \textit{function} of $\nu$, denoted as   $\hat{\mathcal{S}}_\nu$. The \textit{stationarity condition} (Hamilton's principle) requires that
	\begin{equation}\label{E4}
	   \frac{\partial \hat{\mathcal{S}}_\nu}{\partial \nu} \Bigg|_{\nu=1}=0 \ ,
	\end{equation}
	with $\nu$ set to unity in order to recover the initial configuration. The resulting identity is known as the virial identity\footnote{For spherically symmetric configurations, \eqref{E5} can be readily applied to field theory models yielding their virial identities; for other symmetries, one also has to integrate in the angular coordinates.}:
	\begin{equation}\label{E5}
 	 \bigintss_{0}^{+\infty}\left[ \sum_j \frac{\partial \mathcal{L}}{\partial X'_j} X'_j -\mathcal{L} -\frac{\partial \mathcal{L}}{\partial x}x \right]dr = \left[\frac{\partial f}{\partial x}x - \sum_l\frac{\partial f}{\partial X'_l} X'_l\right]^{+\infty}_{0} \ . 
	\end{equation}
    In General Relativity (GR), the Einstein-Hilbert (EH) action contains second-order derivatives of the metric and the total derivative term in~\eqref{E2}, defined by $f$, is non-zero; to obtain the correct virial identity, it is then required to consider the Gibbons-Hawking-York term (GHY)~\cite{york1972role,gibbons1977action} as part of the gravitational action. 

    These results were, however, only shown for the region outside the event horizon in asymptotically flat and Anti-de Sitter spacetimes~\cite{radu2012spinning,herdeiro2015anti,astefanesei2003boson,herdeiro2016static,herdeiro2020class}, where no additional boundary (horizon) exists. The universe is, however, better described in the presence of a positive cosmological constant (de-Sitter), which adds an additional boundary known as the cosmological horizon.
    
    With the assumption that the virial theorem is obeyed for the entire spacetime, although incomplete, the previously discussed methodology indicates the possibility of splitting the spacetime into regions separated by a horizon (region boundaries)~\footnote{While a single virial identity for the whole spacetime seems feasible, the divergence of the metric functions at the horizons makes its use unappealing.}, each with its own identity. From this assumption, two types of regions were identified:
    \begin{itemize}
     \item \textbf{Type A:} In between two given boundaries $r\in [r_-,\, r_+]$\ ,
     \item \textbf{Type B:} Outside the exterior horizon $r\in [r_\text{Ext},\, +\infty[$\ .
    \end{itemize}
    Regions of \textbf{Type A} comprise regions between horizons or from the centre/singularity to a horizon. In contrast, regions of \textbf{Type B} are the well-studied regions outside the external horizon, $r_{Ext}$. 

    While the computation of the virial identity for regions of \textbf{Type B} follows the previously discussed (standard) method, such an approach is impossible in the presence of a second boundary. The existence of an additional boundary forbids  keeping both boundaries invariant under the radial scaling; this is instead allowed  when an asymptotic infinity is present. There is, however, no particular reason for the choice of the radial coordinate presented in \eqref{E1}. One could then conceive a new radial coordinate that permits the same procedure as the standard scaling by keeping one of the boundaries fixed while "sending'' the other boundary to infinity -- effectively making both boundaries invariant under the radial scaling -- allowing the calculation of the virial identity in a bounded region through the usual scaling.
    
    This paper is organised as follows. In Sec.~\ref{S2}, we present the generic method used to generalise the computation of the virial identity for the entire spacetime. The latter is then demonstrated for the Reissner-Nordstrom de-Sitter BH in Sec.~\ref{S2.1} and Sec.~\ref{S2.2}. We finalise in Sec.~\ref{S3} with our conclusions. Throughout this paper, we adopt natural units with $G=1=c$.

%
\section{Method}\label{S2}
%
    As previously mentioned, two types of regions can exist in the presence of an additional boundary. Regions of \textbf{Type B} follow the standard Derrick's scaling argument in the presence of a single boundary: a radial coordinate transformation that translates the horizon to the origin of a new radial coordinate $x\to r-r_\text{Ext}$ and consecutive scaling $x\to \nu x$.
    
    However, regions of \textbf{Type A} require a more elaborated coordinate transformation since an additional boundary exists. In the same spirit as the standard argument, let us consider a transformation that fixes one of the boundaries while sending the other to infinity. Since the boundary at the smallest radius, $r_-$, can include the coordinate origin, for generality, let us consider a coordinate transformation that fixes the outmost boundary, $r_+$, which will always be a horizon
    \begin{equation}\label{E6}
        x\to \frac{r_+-r}{r-r_-}r_+\ ,
    \end{equation}  
    where the multiplication by $r_+$ is used to keep the new coordinate with dimensions of length. While the original radial coordinate could take values of $r\in [r_-,\, r_+]$ in a \textbf{Type A} region, the new one has $x\in[0,\, +\infty[$, effectively making both boundaries invariant under the radial scaling. Computation of the viral identity follows the standard scaling $x\to \nu x$.

\bigskip
    Let us demonstrate our method by applying it to the Reissner-Nordstrom de-Sitter (RNdS) BH solution~\cite{bousso1997charged,antoniadis2020weak,benakli2021dilatonic,chrysostomou2023reissner}. The action describing an electrovacuum BH solution in a de Sitter background comes as  
    \begin{equation}\label{E7}
        \mathcal{S}=\frac{1}{16 \pi}\int \text{d}^4 \textbf{r} \sqrt{-g} \Big[R-2\Lambda-F_{\mu \nu} F^{\mu \nu}\Big]+\frac{1}{8\pi}\int _{\partial \mathcal{M_\pm}} \text{d}^3 \textbf{r}\ \sqrt{h} (\epsilon K - K_0)\ ,
    \end{equation}
    with $g$ the determinant of the metric $g_{\mu \nu}$, $R$ the Ricci scalar, $\Lambda$ a positive cosmological constant, and $F_{\mu \nu}$ the Maxwell tensor given by $F_{\mu \nu}=\partial _\mu A_\nu-\partial _\nu A_\mu$, with $A_\mu = V(r) \text{d}t$ the $4-$vector potential here considered solely electrostatic. Observe that $\Lambda$ is a parameter of the space of theories (a model-dependent degree of freedom) and not a BH parameter\footnote{In the $\Lambda$CDM model, the cosmological constant $\Lambda$ drives the late-type expansion of the universe (\textit{e.g.}~\cite{Planck2018}).}. 
    
\begin{figure}[h!]
     \centering
     \includegraphics[scale=0.85]{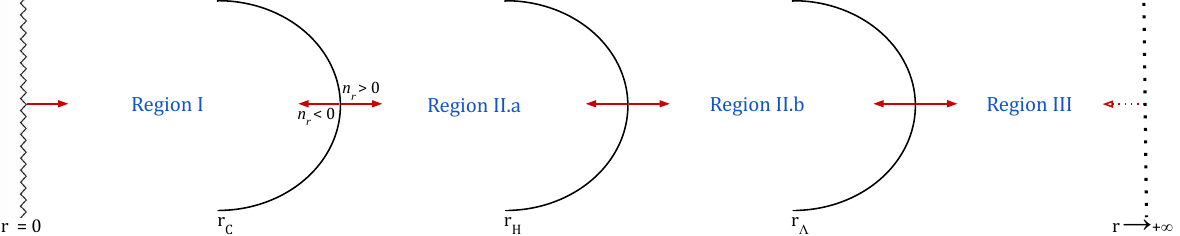}
     \caption{Schematic representation of the Reissner-Nordstrom de Sitter spacetime and its regions.}
     \label{F1}
    \end{figure}

    While the first integral corresponds to the standard Einstein-Hilbert+Maxwell de-Sitter bulk action (thereafter referred by a $b$ index), the second integral is known as GHY term and is necessary in order for the gravitational action to be well-posed. Above, $K= \nabla _\mu n^\mu$ is the extrinsic curvature of the boundary $\partial \mathcal{M}$ with normal $n^\mu$ (graphically represented by a red arrow in Fig.~\ref{F1}), $\epsilon=\pm 1$ depending on whether the normal to $\partial M _\pm$ is spacelike ($+$) or timeline ($-$), and $h$ is the associated induced metric of the boundary. The extra term $K_0$ is the extrinsic curvature of the embedded flat spacetime, a non-dynamical term required for the finite value of the action. 

    In general, the resulting virial identity can be expressed as
    \begin{equation}\label{E8.0}
        \int _0 ^{+\infty} \text{d}x\ I_{b} = I_{GHY}\ ,
    \end{equation}

    where $I_{b}$ represent the contribution from the bulk action and $I_{GHY}$ is the GHY boundary term. For the metric parametrisation, let us consider a generic line element compatible with  spherical symmetry:  
    \begin{equation}\label{E8}
        ds^2 = -H(r)\, dt^2 +\frac{dr^2}{H(r)}+r^2 \big(d\theta ^2 +\sin ^2 \theta\, d \varphi ^2\big)\ .
    \end{equation} 
    Observe that, while the metric $H(r)$ and the electrostatic $V(r)$ functions are kept generic during the computation of the virial identities, they can be expressed analytically in terms of the horizons radii $r_C,\,r_H,\,r_\Lambda$ as
    \begin{equation}\label{E9}
        H(r)=1-\frac{2M}{r}+\frac{Q^2}{r^2}-\frac{\Lambda}{3} r^2 = \frac{(r-r_C) (r-r_H) (r_\Lambda-r) (r+r_C+r_H+r_\Lambda)}{r^2 \Big[r_C^2+r_C (r_H+r_\Lambda)+r_H^2+r_H r_\Lambda+r_\Lambda^2\Big]}\ ,
    \end{equation}
    and $V(r)=-Q/r$, with $M$ the BH's mass and $Q$ it's electric charge. These can also be cast as a function of the horizons radii as:
    \begin{align}\label{E10}
        & M = \frac{(r_C + r_H) (r_C + r_\Lambda) (r_H + r_\Lambda)}{2 \Big[r_C^2 + r_H^2 + r_H r_\Lambda + r_\Lambda^2 + r_C (r_H + r_L)\Big]}\ , \qquad  Q = \frac{\sqrt{r_C} \sqrt{r_H} \sqrt{r_\Lambda} \sqrt{r_C + r_H + r_\Lambda}}{\sqrt{r_C^2+r_H^2 + r_H r_\Lambda + r_\Lambda^2 + r_C (r_H + r_\Lambda)}}\ , \nonumber\\
       & \qquad \qquad \qquad \qquad \qquad \qquad \Lambda = \frac{3}{r_C^2 + r_H^2 + r_H r_\Lambda + r_\Lambda^2 + r_C (r_H + r_\Lambda)} \ .
    \end{align} 
    Regularity of the horizons imposes a bound on $M^2 \Lambda$ \cite{romans1992supersymmetric,bousso1997charged}
    \begin{equation}\label{E11}
      M^2 \Lambda \leqslant \frac{1}{18}\Big[1+12 Q^2 \Lambda+\big( 1-4 Q^2 \Lambda \big)^{\frac{3}{2}} \Big]\ .
    \end{equation}

    The RNdS spacetime has three horizons, $0<r_C<r_H<r_\Lambda$ (see Fig.~\ref{F1}), which divide the spacetime into a region of \textbf{Type B} -- from the outermost horizon, ($r_\text{Ext}=r_\Lambda\equiv$ Cosmological horizon) to asymptotically infinity -- and three regions of \textbf{Type A}: \textit{Region I} spans from the singularity until the Cauchy horizon, $r\in [0,\, r_C]$, with $r_-=0$ and $r_+=r_C$. \textit{Region II.a} covers the region in between the Cauchy and the Event horizon: $r\in [r_C,\, r_H]$, with $r_-=r_C$ and $r_+= r_H$. Finally, \textit{Region II.b} covers the region in between the Event and the Cosmological horizon: $r\in [r_H,\, r_\Lambda]$, with $r_-=r_H$ and $r_+=r_\Lambda$.

    Due to the absence of angular coordinates dependencies, the reduced Lagrangian and total derivative are
    \begin{equation}
        \mathcal{L}= -2 \Big[-1+H+r^2\Lambda + r\big( H'-rV'\big)\Big]\ , \qquad \qquad {\rm and} \qquad \qquad f=-r^2 H'\ .
    \end{equation}
%
    \subsection{Type A}\label{S2.1}
%
    Let us start by computing the virial identity in the regions between two boundaries. From the coordinate transformation \eqref{E6}, the boundary $r_+$ is now localised at $x=0$, while the second boundary $r_-$ goes to infinity, $x\to +\infty$. Due to the coordinate transformation, one must also transform the metric and matter functions accordingly, $X(r)\to X(x)$, as well as all associated derivatives. The resulting reduced Lagrangian comes as
    \begin{align}
        \mathcal{L}^A =\left[\frac{(r_+ +x) \left(xr_- +r_+^2\right) \Big((r_++x) \left(xr_-+r_+^2\right) V'^2-r_+ (r_--r_+) H'\Big)}{r_+^2 (r_--r_+)^2}-H-\frac{\Lambda  \left(xr_- +r_+^2\right)^2}{(r_++x)^2}+1\right]\ .
    \end{align}
    The computation of the viral identity follows the standard Derrick's scaling argument in the absence of a horizon: $x\to \nu x$ and $X^{(n)}(\nu x)\to \frac{X_\nu ^{(n)} (x)}{\nu ^n}$, with $n$ the derivation order. Following \eqref{E5} and \eqref{E8.0}, one gets
    \begin{align}
        &I_{b} ^A= \Bigg[ \frac{xr_+ (r_+-r_-) H'}{(r_+ +x)^2}+\frac{r_+ H (r_+-r_-) (r_+-x)}{(r_++x)^3}+\frac{\left(x^2 r_-^2-r_+^4\right) V'^2}{r_+ (r_--r_+)}\nonumber\\
        &+\frac{r_+ (r_--r_+) \Big((r_+-x) (r_++x)^2-\Lambda  \left(x r_-+r_+^2\right) \big(3r_+ x (r_--r_+)-x^2 r_- +r_+^3\big)\Big)}{(r_++x)^5}\Bigg]\ .
    \end{align}
    The contribution of the bulk action to the virial identity can then be obtained by replacing the transformed metric and electrostatic functions, $H(x)$ and $V(x)$, and subsequent integration of the new radial coordinate $x$:
    \begin{align}
     \bigintsss _0 ^\infty dx\ I_{b}^A =\frac{r_+^5 x (r_--r_+)(\cdots)}{(r_++x)^4 \left(xr_-+r_+^2\right)^2}+\cdots +\frac{2 (r_+-r_-) x^5 (\cdots)}{(r_++x)^4 \left(x r_- +r_+^2\right)^2}\Bigg|_{0} ^{\infty} =0 \ ,
    \end{align}
    where we showed only the leading contribution to lowest and highest order terms in $x$. 
    
    Remains now to compute the GHY boundary term for regions of \textbf{Type A}. The normal vector is given by $n_r=\sqrt{H}\, \partial _r$; the respective induced metric determinant, $\sqrt{-h}$, and extrinsic curvatures at the boundaries, $K$, are
    \begin{equation}
     \sqrt{-h}=\frac{(r_--r_+)(r_++xr_-)^2 \sqrt{H}}{(1+x)^4}\ , \qquad K=\frac{(r_++x) \Big[ \frac{(r_++x) H' \left(xr_-+r_+^2\right)}{r_+ (r_--r_+)}+4 H\Big] }{2 \sqrt{H}\big(xr_-+r_+^2\big)}\ ,
    \end{equation}
    with the extrinsic curvature of the embedded flat spacetime $K_0=-\frac{2 (x+1)}{xr_-+r_+}$. The associated total derivative from the GHY term is then
    \begin{equation}
        f_{GHY} ^A =\left( xr_-+r_+^2\right) \left(\frac{4 (H-1)}{r_++x}-\frac{H' \left(xr_-+r_+^2\right)}{(r_+-r_-)r_+}\right)\ .
    \end{equation}
    Observe that the GHY term completely cancels the total derivative coming from the effective Lagrangian, $f=-\frac{H' \left(xr_-+r_+^2\right)^2}{(r_--r_+)r_+}$, effectively removing the second-order derivatives from the complete effective action of our model (precisely the goal of the boundary term). The remaining contribution from the GHY term to the virial identity is then
    \begin{equation}
        I_{GHY} ^A= \frac{4 (r_--r_+)  (H-1) xr_+}{(r_++x)^2}\Bigg|_0 ^\infty =0\ .
    \end{equation}
    Interestingly, while the convenient parametrisation presented in \cite{herdeiro2021virial}, $H(r)=1-\frac{2m(r)}{r}$, \textit{does not} trivialise the gravitational contribution to the virial identity -- due to the cosmological constant contribution to the action -- the more generic parametrisation of the metric, $H(r)$, together with the radial transformation \eqref{E6}, does. This allows the computation of the virial identity solely associated with the matter components of the action under study, extremely simplifying the computation and analysis.
%
    \subsection{Type B}\label{S2.2}
%
   At last, consider the region outside the external horizon, \textit{Region III} in Fig.~\ref{F1}. In this region, the radial transformation follows the traditional radial translation in the presence of a horizon~\cite{herdeiro2021virial,herdeiro2022deconstructing,oliveira2023convenient}, \textbf{Type B}: $ x\to r-r_{Ext}$, with $r_{Ext}\equiv r_\Lambda$.  
      After performing the appropriated coordinate transformation, $X(r)\to X(x)$, and respective derivatives, the resulting reduced Lagrangian comes as
    \begin{equation}
        \mathcal{L}^{B}  =\Big[\big( r_\Lambda+x\big) \Big(\big(r_\Lambda+x\big) V'^2-H'\Big)-H-\Lambda  \big(r_\Lambda+x\big)^2+1\Big]\ .
    \end{equation}
   Following the standard procedure \eqref{E5}, the resulting contribution of the bulk action to the virial identity is
    \begin{align}
    \bigintsss _0 ^{+\infty} I_b ^B = & -\frac{\Lambda r_C r_\lambda}{3 (r_\Lambda+x)^2 } \Bigg[\big( r_C+r_H+r_\Lambda\big)\Big( x \big( 3 r_H+r_\Lambda\big)+r_\Lambda \big(r_H+r_\Lambda\big)\Big)\nonumber\\
    & \qquad \quad+\frac{\big(r_\Lambda+x\big)}{r_C r_\Lambda} \Big(r_H^2 r_\Lambda^2+r_H r_\Lambda^3+2 x \big( r_\Lambda+x\big)^3\Big)\Bigg]_0 ^{+\infty}\nonumber\\
    & = 2 M+\frac{\Lambda}{3}   \Big[-2 r_\Lambda^2 x-4 r_\Lambda x^2-2 x^3\Big]_{x\to +\infty}\ .
    \end{align}
    One can see that the divergence at spatial infinity is associated with the spacetime's de-Sitter nature, $\Lambda\neq 0$. Computing now the GHY boundary term, we have that the total derivative $f$ from the bulk action is, again, completely cancelled, resulting in an effective contribution to the virial identity, $f^B$:
    \begin{equation}
        f_{GHY} ^B =(r_\Lambda+x) \left[(r_\Lambda+x) H'+2 H-2\right] \ ,\qquad \qquad f^B=2 \big( H-1\big) \big(r_\Lambda+x\big)\ .
    \end{equation}
    The remaining contribution from the GHY term to the virial identity is
    \begin{equation}
        I_{GHY} ^B= 2 x H\Big|_0 ^\infty =2 M+\frac{\Lambda}{3}   \Big[-2 r_\Lambda^2 x-4 r_\Lambda x^2-2 x^3\Big]_{x\to +\infty}\ ,
    \end{equation}
    completely cancelling the contribution of the bulk action to the virial identity and associated divergence.
%
\section{Conclusion}\label{S3}
%
    In this work, we have introduced a method based on a radial coordinate transformation that generalises Derrick's scaling argument to a region of the spacetime bounded by either two horizons or one horizon and a coordinate origin/singularity (here referred to as \textbf{Type A}). While in the presence of a sole boundary (\textbf{Type B} region), a single coordinate translation is performed before the scaling, in the presence of a second boundary, such is not possible.

    The problem lies in the impossibility of scaling the radial coordinate without changing the boundary. In a region with a single boundary, fixing the latter while scaling the radius does not interfere with the asymptotic behaviour of the newly scaled radial coordinate. However, such is unfeasible in the presence of a second boundary. A more evolved radial transformation is necessary. In this work, we suggest a coordinate transformation that fixes one of the boundaries while sending the other to infinity, effectively making both boundaries invariant under the scaling.

    The method was tested using a Reissner-Nordstrom de-Sitter BH solution. The latter contains three regions delimited by two boundaries and one region that goes to radial infinity. In the three regions of \textbf{Type A}, the proposed radial transformation resulted in a trivial contribution of the gravitational action to the virial identity.

    In the case of the \textbf{Type B} region, while the usual Derrick scaling is valid, the cosmological constant factor introduces a divergency of the bulk contribution to the virial identity. Fortunately, such divergence was entirely quenched by the contribution of the boundary term (GHY term), verifying the identity in the RNdS spacetime.

    While the current results are only presented for spherical symmetry, one expects the same procedure to be valid for axial symmetry with the added complexity of the angular dependence.

    At last, while we did not establish any new no-hair/go theorem with the aid of the new method and metric parametrisation presented, one expects that it can seed new theorems – just like it was done in spherical symmetry.
%
\section*{Acknowledgments}
%
 We would like to thank Eugen Radu for his valuable discussions and comments. A. M. Pombo is supported by the Czech Grant Agency (GA\^CR) under grant number 21-16583M.
 
\bibliographystyle{ieeetr} 
\bibliography{master}

\end{document}